%% file: main_arxiv.tex
\documentclass[conference]{IEEEtran}

\IEEEoverridecommandlockouts

\usepackage{cite}
\usepackage{xcolor}
\usepackage{stfloats}
\usepackage[caption=false,font=footnotesize]{subfig}
\usepackage{amsmath,amssymb,amsfonts}
\usepackage{mathtools}
\usepackage{algorithm}
\usepackage{algorithmic}
\usepackage{graphicx}
\usepackage{textcomp}
\usepackage{xcolor}
\usepackage{comment}
\usepackage{lipsum}
\usepackage{footmisc}
\usepackage{tikz}
\usetikzlibrary{circuits.logic.US}
\usetikzlibrary{positioning}
\usetikzlibrary{decorations,bending,decorations.pathreplacing,arrows.meta,spath3}
\usepackage{xparse}
\usepackage{pgfplots}
\pgfplotsset{compat=1.18}
\makeatletter
\tikzset{%
  get path length/.code={%
    \tikz@addoption{%
      \pgfgetpath\tikz@tmppath%
      \pgfprocessround\tikz@tmppath\tikz@tmppath%
      \pgf@decorate@parsesoftpath\tikz@tmppath\tikz@discard%
      \global\let#1=\pgf@decorate@totalpathlength%
    }%
  }
}
\makeatother

\NewDocumentCommand\drawarrow{omm}{
  \path[cyan,get path length=\a,  spath/save=arrow] (#2)  to[#1]  (#3);
  \path[cyan,shorten >=0.27*\a, spath/use=arrow, spath/save=path];
  \draw[cyan,-{Stealth[cyan,
    length=0.8cm,
    bend]},spath/use=arrow];
  \draw[cyan,spath/use=path,
    decoration={expanding waves,
        angle=5,
        segment length=0.1},{decorate}];
  }
\usepackage{hyperref}
\usepackage{pgfplotstable}
\usetikzlibrary{spy}
\usepackage{comment}
\usepackage{soul}
\setstcolor{red}
\setulcolor{blue}
\setul{0.5ex}{1.5pt}

\usepackage{booktabs}
\usepackage{siunitx}
\usepackage{threeparttable}
\usepackage[mode=buildnew]{standalone}
\usepackage[table]{xcolor}
\usetikzlibrary {arrows.meta,graphs,shapes.misc}
\usetikzlibrary{shapes.geometric}
\usetikzlibrary{positioning, calc, decorations.pathreplacing}
\usetikzlibrary{circuits.logic.US,tikzmark}
\usetikzlibrary{shapes}
\usetikzlibrary{shadows.blur}
\usetikzlibrary{positioning}
\usetikzlibrary{plotmarks}
\usetikzlibrary{shapes.geometric, arrows.meta, positioning, calc, fit, backgrounds}

    \def\BibTeX{{\rm B\kern-.05em{\sc i\kern-.025em b}\kern-.08em
    T\kern-.1667em\lower.7ex\hbox{E}\kern-.125emX}}
    
\hypersetup{          
    colorlinks=true,
    linkcolor=black,
    filecolor=magenta,      
    urlcolor=cyan,
    citecolor=teal!80,
    pdftitle={},
}
\definecolor{myblue}{RGB}{0,114,178}
\definecolor{myred}{RGB}{213,94,0}
\definecolor{mygreen}{RGB}{0,158,115}    
\definecolor{mygray}{RGB}{70,70,70}
\definecolor{lightgrid}{RGB}{225,225,225}
\definecolor{mycyan}{RGB}{0,150,170} 
\definecolor{mydarkgray}{RGB}{45,45,45} 
\definecolor{mypurple}{RGB}{117,112,179}

\usepackage{balance}

\def\paperdoi{}
\usepackage{eso-pic}
\newcommand{\ieeenotice}{%
  \parbox{\textwidth}{\scriptsize\centering
  This work has been accepted for publication in the 2026 IEEE 33rd International
  Conference on Electronics, Circuits and Systems (ICECS).%
  \ifx\paperdoi\empty\else\ DOI: \href{https://doi.org/\paperdoi}{\paperdoi}.\fi\\
  \copyright~2026 IEEE. Personal use of this material is permitted. Permission from IEEE
  must be obtained for all other uses, in any current or future media, including
  reprinting/republishing this material for advertising or promotional purposes,
  creating new collective works, for resale or redistribution to servers or lists,
  or reuse of any copyrighted component of this work in other works.}}
\AddToShipoutPictureFG*{%
  \AtPageLowerLeft{\put(\LenToUnit{0.5\paperwidth},\LenToUnit{0.32in}){%
    \makebox[0pt][c]{\ieeenotice}}}}
  
\begin{document}


\title{Core-Lock Scheduling for 5G NR LDPC Decoders}


%
\author{%
\IEEEauthorblockN{%
N.~Papageorgiou and V.~Paliouras
}
\IEEEauthorblockA{Electrical and Computer Engineering Department, University of Patras, Greece
}}
%

\maketitle

\begin{abstract} 
Accelerating iterative LDPC decoding while preserving error-correction reliability is a critical challenge in latency-constrained 5G New Radio (NR) and non-terrestrial networks. This paper proposes a hardware-efficient core-lock scheduling mechanism that evaluates the core syndrome weight after each layer update to selectively bypass extension layers. With the selected threshold $\theta{=}0.20$, the mechanism reduces decoding latency by up to 31.3\% when integrated with static schedules, and by up to 45.5\% compared to alternative dynamic schemes utilizing the full parity-check matrix, without block error rate (BLER) degradation. Compared with an estimated fully parallel full-matrix checker, the core-only approach yields a 76\% reduction in shifter network area and a 73\% reduction in XOR gate count. For decoders with existing early termination, the proposed mechanism requires only a compact adder tree and a comparator.
\end{abstract}

\vspace*{-0.15cm}

\vspace{-0.2cm}
\section{Introduction}
\label{s:intro}
\vspace{-0.1cm}
Modern wireless systems, including 5G New Radio (NR) and non-terrestrial networks (NTNs), impose stringent throughput and latency demands on channel decoders~\cite{9711564}. In NTN scenarios, long propagation delays limit conventional hybrid automatic repeat request (HARQ), increasing the importance of first-transmission decoding success~\cite{10975810}. These, motivate decoding techniques that accelerate convergence while maintaining reliability and low hardware complexity.

Low-density parity-check (LDPC) codes~\cite{1057683} have been adopted as the standard channel coding scheme for the enhanced mobile broadband (eMBB) of 5G NR due to their near capacity performance~\cite{etsi_ts_138212_v17_5_0}. 
Iterative decoding typically employs belief propagation (BP)~\cite{910572}, whereas hardware implementations favor the lower-complexity min-sum (MS) approximation~\cite{768759} and its offset (OMS) and normalized (NMS) variants~\cite{1001666}. Beyond simplifying message computations, optimizing their update order can further accelerate convergence.

Among dynamic scheduling methods that adapt the update sequence during decoding, informed dynamic scheduling (IDS)~\cite{5610969} prioritizes variable-node updates based on residual errors, a concept from which several subsequent variations have been developed~\cite{8664167}. However, these dynamic approaches require intensive online computations, motivating the development of hardware-friendly static layered schedules. Bottom-to-top (BtT) scheduling~\cite{8661345} prioritizes lower-degree layers to improve early information exchange, while base-graph-based static scheduling (BGSS)~\cite{9765969} also considers punctured nodes and row weights. Static rescheduling using localized quantized reliability metrics was proposed in~\cite{papageorgiou2025novel}. Additionally~\cite{9222014} shows that extension rows contribute minimally after reaching a convergence margin, allowing their omission in later iterations. Dynamic transitions between core and extension processing consider this limitation~\cite{11154636}, but introduce additional online metric computations and control logic.

In this paper, we propose a hardware-efficient core-lock scheduling mechanism for 5G NR LDPC decoders. A lifting-size-normalized core syndrome weight is evaluated after each layer update to trigger a transition to core-only processing, bypassing extension layers. The mechanism can augment any underlying static schedule. We analyze the threshold trade-off between latency and block error rate (BLER) and evaluate the mechanism with established static schedules and a simplified dynamic scheduler. To assess its hardware cost, we present fully parallel and time-multiplexed implementations with reuse of existing early-termination circuitry where available.

The remainder of this paper is organized as follows. Section~\ref{pre} provides preliminaries on 5G NR LDPC codes and layered decoding. Section~\ref{pro} details the proposed continuous core-lock mechanism and dynamic thresholding logic, while Section~\ref{eval} presents the threshold trade-off analysis and a comprehensive algorithmic evaluation. Section~\ref{hw} discusses the hardware implementation and ASIC synthesis results, while conclusions are drawn in Section~\ref{con}.

\vspace{-0.15cm}
\section{Preliminaries}
\label{pre}
\vspace{-0.05cm}
We consider the concatenated structure defined in the 5G NR standard~\cite{etsi_ts_138212_v17_5_0}, where an $A$-bit information sequence is first encoded by an outer CRC and subsequently processed by an inner quasi-cyclic LDPC encoder producing a codeword $\mathbf{c} {\in} \{0,1\}^N$. The parity-check matrix $\mathbf{H}_{M\times N}$ is constructed by expanding one of two standardized base graphs, BG1$_{46\times 68}$ or BG2$_{42\times 52}$, by a lifting size $Z_c {\in} \{1, \dots, 384\}$~\cite{etsi_ts_138212_v17_5_0}. Iterative decoding of LDPC codes is typically performed using a row-layered scheduling strategy to accelerate convergence~\cite{1363033}. Under this approach, the matrix $\mathbf{H}$ is partitioned into horizontal layers, each corresponding to a row of the initial base graph. The core submatrix $\mathbf{H_c}$ is obtained by lifting the first four base-graph rows and the first 26 columns for BG1 or 14 columns for BG2.  In our evaluations, codeword validation is thus simplified to verifying the core syndrome relation $\mathbf{s}_c{=}\mathbf{\hat{c}}_c \mathbf{H}_c^{\mathrm{T}}{=}\mathbf{0} \pmod 2$~\cite{papageorgiou2025novel}, where $\mathbf{\hat{c}}_c$ represents the sub-vector of the estimated $\mathbf{c}$, corresponding to the core columns.

\vspace{-0.1cm}
\section{Proposed Dynamic Core-Lock Scheduling}
\label{pro}

\input{theta}

Although advanced static schedules accelerate decoding convergence, their fixed processing order does not adapt to the evolving decoder state. As information propagates from the extension layers to the core variable nodes, further extension-layer updates may provide diminishing gains while consuming additional clock cycles. We therefore propose a core-lock mechanism that augments a baseline static schedule by switching to core-only processing at a sub-iteration level.

To dynamically track decoding convergence without the hardware cost of full-matrix syndrome evaluations~\cite{11154636}, the proposed architecture monitors exclusively the number of unsatisfied core parity checks through the syndrome vector $\mathbf{s}_c$. Since the core spans four base-graph rows, its lifted parity-check matrix $\mathbf{H}_c$ contains $4Z_c$ check equations. The core syndrome weight $W_{\mathrm{core}}$ is defined as 
\begin{equation}
    W_{\mathrm{core}}=\|\mathbf{s}_c\|_0 = \textstyle\sum\nolimits_{i=1}^{4 Z_c} s_{c,i},
    \label{eq:core_weight}
\end{equation}
where $s_{c,i} \in \{0, 1\}$ is the $i$th element of $\mathbf{s}_c$. To ensure that the decoder-state transition criterion remains universal and independent of $Z_c$, we define the normalized syndrome weight 
\begin{equation}
    W_{\mathrm{norm}} = W_{\mathrm{core}} / Z_c.
    \label{eq:normalized_weight}
\end{equation}
Since $W_{\mathrm{norm}}\in[0,4]$, a threshold $\theta\in[0,4]$ defines the core-lock condition 
\begin{equation}
    W_{\mathrm{norm}} \leq \theta.
    \label{eq:threshold_condition}
\end{equation}
The criterion is evaluated after each layer update, allowing a scheduling transition within an iteration. Once the threshold condition~\eqref{eq:threshold_condition} is satisfied, convergence of the systematic core region is inferred, and the remaining extension-layer updates are bypassed for the rest of decoding process.

\input{evalplot}
\vspace*{-0.1cm}
\section{Algorithmic Evaluation and Results}
\label{eval}

We evaluate the proposed mechanism using 5G NR LDPC codes over an AWGN channel with QPSK modulation and a layered NMS decoder limited to 20 iterations. Core-only ET is applied after each layer update~\cite{papageorgiou2025novel}. We first examine the trade-off between decoding latency and BLER to select an operating threshold $\theta$, and then benchmark the proposed architecture against existing solutions~\cite{8661345,9765969,11154636}.

\vspace{-0.1cm}
\subsection{Core-Lock Threshold Extraction and Trade-off Analysis}
The operating threshold $\theta$ is selected by assessing the trade-off between BLER and decoding latency, represented by the average number of iterations (ANI), for $\theta\in\{0.20,0.40,0.60\}$. Fig.~\ref{thetasweep} compares these thresholds with the sequential baseline without core-lock for BG1 with $A{=}4200$ and $R{=}1/3$.

An increase in $\theta$ yields greater decoding speedup at the expense of BLER degradation. As shown in Fig.~\ref{thetasweep}(a),  $\theta{=}0.40$ exhibits BLER degradation in the $10^{-4}$ region,  whereas the baseline successfully converges. This degradation is further exacerbated under the highly aggressive threshold of $\theta{=}0.60$, which exhibits an apparent error floor near $5\times10^{-2}$. This limitation is attributed to the structure of the 5G NR LDPC code, which is a concatenation of a core code and a Low-Density Generator Matrix (LDGM) extension~\cite{8316763}. Under high thresholds, the core-lock trigger is satisfied prematurely, deactivating the extension check nodes before the extrinsic information from the LDGM part has propagated sufficiently to assist the core in clearing residual errors. Conversely, $\theta{=}0.20$ closely follows the baseline BLER over the evaluated SNR range for this BG1 configuration.

Figs.~\ref{thetasweep}(b) and~(c) show latency savings of $6.6\%$--$33.1\%$ for $\theta{=}0.20$, corresponding to $0.5$--$1.2$ fewer iterations on average, depending on channel conditions. Although more aggressive thresholds provide latency savings of up to 35.7\%, their BLER penalty motivates selecting $\theta{=}0.20$ which is also used in the subsequent evaluation. The core-lock criterion applies to both BG and different rates, with optional configuration-dependent threshold to adjust the trade-off between latency and BLER.

\vspace*{-0.1cm}
\subsection{Comparative Latency and Error-Correction Performance}
To benchmark the proposed core-lock mechanism, we integrate it into the topology-aware BGSS~\cite{9765969} and BtT~\cite{8661345} static schedules. We further compare and evaluate it within the dynamic multi-state architecture of~\cite{11154636}, replacing its original criteria and bypassing the extension-only (EO)~state.

As shown in Fig.~\ref{maineval}(a), applying the core-lock with $\theta{=}0.20$ to the static schedules preserves decoding reliability, introducing no error-correction penalty while yielding a slight BLER improvement across the waterfall region. The corresponding ANI reduction in Fig.~\ref{maineval}(b) ranges from $0.2$ to $1.2$ iterations per frame, translating into latency savings of $10.7\%$--$29.1\%$ for BGSS and $10.8\%$--$31.3\%$ for BtT, as illustrated in Fig.~\ref{maineval}(c).

For the dynamic architecture of~\cite{11154636}, replacing its original $15\%$ and $35\%$ full-matrix thresholds, used without independent retuning, with core-only thresholds $\theta_{\text{CO}}{=}0.20$, $\theta_{\text{ECC}}{=}0.60$ maintains comparable BLER, with a small ANI increase at low SNR. This substitution eliminates full-matrix syndrome evaluation, reducing hardware requirements, and further calibration of  $\theta_{\text{ECC}}$ could close this latency gap. Critically, the EO~state of~\cite{11154636} relies on calculating the mean absolute LLR magnitude, a hardware-expensive operation requiring wide adder trees and division. Bypassing this state simplifies the architecture but produces pronounced error floors with the original full-$\mathbf{H}$ thresholds. In contrast, the core-only metric maintains decoding reliability over the evaluated SNR range, as shown in Fig.~\ref{maineval}(d). With EO bypassed, the proposed scheme achieves latency reductions of $3.8\%$--$45.5\%$ relative to the original dynamic baseline  with EO enabled~\cite{11154636}, as shown in Figs.~\ref{maineval}(e) and~(f). Additional evaluations cover $A{=}502$ at $R\in\{1/5,1/2\}$ for BG2 and $A{=}3848$ at $R{=}1/2$ for BG1. Across the verified scheduling configurations, the core-only criterion with $\theta{=}0.20$ exhibits results consistent with Fig.~\ref{maineval}, supporting its use across different block lengths and code rates.

\vspace*{-0.21cm}
\section{Hardware Architecture}
\label{hw}
\vspace*{-0.05cm}
We evaluate the proposed architecture using a Cadence synthesis flow and a commercial 28-nm standard-cell library. Hardware complexity is compared with an estimated fully parallel full-$\mathbf{H}$ syndrome checker for the metric in~\cite{11154636}. The proposed mechanism, shown in Fig.~\ref{fig:HW}, is implemented as a combinational feed-forward pipeline operating on the $\mathbf{H}_c$, supporting the worst-case $Z_{\max}{=}384$. In the first stage, hard-decision bits are cyclically shifted by 19 parallel 384-bit reconfigurable barrel shifters (BR)~\cite{papageorgiou2025novel, 8052499} per core row. Each BR occupies $3{,}921$~\textmu{}m$^2$, with a delay of $1.00\,\mathrm{ns}$ and power consumption of $0.24\,\mathrm{mW}$. Using the component results in Table~\ref{tab:synthesis_results}, the 76 BRs core network occupies approximately $0.30\,\mathrm{mm}^2$, compared with $1.24\,\mathrm{mm}^2$ for 316 BRs in the full-matrix checker, a 76\% reduction. The aligned bits feed 384 instances of XOR trees per core row, each containing 18 two-input XOR gates. The four rows therefore require 27,648 XOR gates, 73\% fewer than the 103,680 gates in the full-matrix realization.  Each component of 384 XOR-tree instances in Fig.~\ref{fig:HW}, exhibits a delay of $0.16$\,ns and a power consumption of $0.30$\,mW. The resulting 1,536 core syndrome bits are then accumulated using a combinational population-count tree. Since scheduling requires only a threshold comparison, partial sums can saturate above the largest supported threshold without changing the decision. This bounds the upper-stage arithmetic to seven bits, covering the worst-case threshold $0.20{\cdot}384{\approx}77$, instead of the eleven bits required for an exact core syndrome count. The adder tree introduces a delay of $1.58$\,ns and consumes $0.40$\,mW. The dedicated logic occupies $3{,}539$\,\textmu{}m$^2$, whereas a full-matrix evaluation over 17,664 syndrome bits would increase the popcount overhead by at least  11$\times$. The saturated count ($W_{\mathrm{core}}$) is compared with $T_{\mathrm{limit}}{=}\theta Z_c$, stored in a $51{\times}7$ ROM of precomputed 5G NR lifting sizes, and the comparator output drives  the index extraction unit where a multiplexer selects the next processed layer. The listed BR, XOR, and population-count blocks together consume $19.84\,\mathrm{mW}$. Their delays sum to $2.74\,\mathrm{ns}$, below the approximately $3.00\,\mathrm{ns}$ period of a $333\,\text{MHz}$ clock. Higher clock frequencies of the LDPC decoder may require pipeline registers at the cost of an additional clock cycle and resulting to deferring the scheduling decision to the subsequent layer. Crucially, in the common case where the baseline decoder already features standard ET, the BRs and XOR trees can be fully reused. The additional circuit then comprises the population-count tree, threshold lookup, comparison, and scheduling control. Bypassing the EO state also removes the separate mean-LLR accumulation and normalization logic required by~\cite{11154636}.
\begin{figure}[t]
    \centering
    \input{FigArch}
    \caption{Architecture overview of the core-lock mechanism.
    }
    \vspace{-0.6cm}
    \label{fig:HW}
\end{figure}
For heavily area-constrained designs lacking ET hardware that cannot support the fully parallel architecture, the core-lock checker can be time multiplexed. Using only two or one evaluation units reduces the dedicated hardware by 50\% or 75\%, respectively. Hardware multiplexing introduces a temporal trade-off, illustrated in the timing diagram of Fig.~\ref{fig:delay}. To ensure algorithmic consistency, the proposed mechanism process a snapshot of the LLR signs despite possible changes in intermediate stages. Because the main decoder datapath continues during the sequential evaluation (e.g., 4 cycles for a 1-unit design), a delay occurs in the realization and implementation of the scheduling changes. In the worst-case scenario, shown in Fig.~\ref{fig:delay}, if the transition condition is met immediately after a snapshot is taken (i.e., at the end of processing layer A), the architecture only registers it in the next snapshot (captured at layer E) and applies the core-lock schedule change (B$^\prime$) at layer I. This results in a 7-layer speculative pipeline overrun per decision, and up to 14 layers of extra processing if both the schedule-switch and final codeword termination (detected at P) are delayed. In contrast to the fully parallel implementation, which incurs no scheduling latency, the time-multiplexed architecture trades modest scheduling latency for significant hardware savings.

\vspace*{-0.15cm}
\section{Conclusion}
\label{con}
\vspace*{-0.05cm}
This paper presents a hardware-efficient core-lock scheduling mechanism for 5G NR LDPC decoders. Monitoring the core syndrome weight after each layer update enables extension-layer bypassing without requiring computationally intensive state metrics. Algorithmic evaluations show that with a selected threshold of $\theta{=}0.20$ decoding latency is reduced by 10.7\%--31.3\% across static schedules, and by up to 45.5\% compared to a similar dynamic scheduler utilizing the full-$\mathbf{H}$ matrix~\cite{11154636}, with zero BLER degradation. The core-only design reduces shifting network area by 76\% and XOR gate count by 73\% compared to a conventional full-$\mathbf{H}$ checker, requiring only $3{,}539$~\textmu{}m$^2$ in a 28-nm technology when early termination is already present. These results confirm the core-lock mechanism as a practical, low-overhead solution for high-throughput decoders.

\input{timing}

\begin{table}[t]
\centering
\caption{Post-Synthesis Results at 28~\upshape{nm}}
\label{tab:synthesis_results}
\small
\resizebox{\columnwidth}{!}{%
\begin{tabular}{lccc}
\toprule
\textbf{Component} &  \textbf{Area (\textmu m$^2$)} & \textbf{Critical Path} & \textbf{Power}\\ 
\midrule
384-bit BR & 3921 & 1~ns & 0.24~mW\\
384~inst. XOR Trees & 5765 & 0.16~ns & 0.3~mW\\
Saturating Popcount Adder Tree  & 3539 & 1.58~ns & 0.4~mW\\ 
\bottomrule
\end{tabular}}
\end{table}

\bibliographystyle{IEEEtran}
\bibliography{IEEEabrv,bibl}

\end{document}

%% file: theta.tex
\begin{figure*}[t]
    \centering

    \hspace{-0.3cm}
    \subfloat{
        \begin{tikzpicture}[baseline]
            \begin{semilogyaxis}[
            scale only axis,
                width=5.5cm,
                height=3.7cm,
                xmin=-1.45, xmax=0.3,     
                ymin=1e-5, ymax=1,       
                xlabel={\footnotesize{SNR (dB)} \\[-0.5ex] \footnotesize{(a)}}, 
                xlabel style={yshift=1mm, align=center},
                ylabel={\footnotesize {BLER}},
                x tick label style={font=\footnotesize},
                ylabel style={yshift=-1.5mm},
                y tick label style={rotate=90, font=\footnotesize}, 
                grid=both,
                minor tick num=1,
                legend pos=south west,
                legend style={
                    row sep=0.02cm,
                    font=\footnotesize,
                    cells={align=left},
                    legend cell align={left},
                    nodes={scale=0.85, transform shape},
                    at={(0.001,0.001)}
                },]
                
            \addplot[black, mark=square, line width=1.2pt, mark size=1.7pt]
                table[x=SNR, y=BLER_Base]{ThetaSweep.txt};
            \addlegendentry{Baseline}
            
            \addplot[myblue!80, mark=triangle, line width=1.2pt, mark size=1.7pt] 
                table[x=SNR, y=BLER_T20]{ThetaSweep.txt};
            \addlegendentry{$\theta = 0.20$}

            \addplot[myred!70, mark=x, line width=1.2pt, mark size=2.0pt]  
                table[x=SNR, y=BLER_T40]{ThetaSweep.txt};
            \addlegendentry{$\theta = 0.40$}

            \addplot[mygray!60, mark=diamond, line width=1.2pt, mark size=2.0pt]  
                table[x=SNR, y=BLER_T60]{ThetaSweep.txt};
            \addlegendentry{$\theta = 0.60$}
\end{semilogyaxis}
\end{tikzpicture}
    }
    \hspace{-0.42cm}
    \subfloat{
        \begin{tikzpicture}[baseline=(current axis.south)]
            \begin{axis}[
            scale only axis,
                width=5.3cm,
                height=3.7cm,
                xmin=-1, xmax=3.3,
                ymin=0, ymax=15,        
                xlabel={\footnotesize{SNR (dB)}\\[-0.5ex] \footnotesize{(b)}}, 
                xlabel style={yshift=1mm, align=center},
                ylabel={\footnotesize {Average Number of Iterations}},
                x tick label style={font=\footnotesize},
                ylabel style={yshift=-1mm},
                y tick label style={rotate=90, font=\footnotesize}, 
                grid=both,
                minor tick num=1,
                legend pos=north east,
                legend style={
                    row sep=0.02cm,
                    font=\footnotesize,
                    cells={align=left},
                    legend cell align={left},
                    nodes={scale=0.85, transform shape},
                    at={(0.999,0.999)}
                },]
            
            \addplot[black, mark=square, line width=1.2pt, mark size=1.7pt]
                table[x=SNR, y=ANI_Base]{ThetaSweep.txt};
            \addlegendentry{Baseline}    
                
            \addplot[myblue!80, mark=triangle, line width=1.2pt, mark size=1.7pt] 
                table[x=SNR, y=ANI_T20]{ThetaSweep.txt};
             \addlegendentry{$\theta = 0.20$}   

            \addplot[myred, mark=x, line width=1.2pt, mark size=2.0pt] 
                table[x=SNR, y=ANI_T40]{ThetaSweep.txt};
            \addlegendentry{$\theta = 0.40$}

            \addplot[mygray!60, mark=diamond, line width=1.2pt, mark size=2.0pt] 
                table[x=SNR, y=ANI_T60]{ThetaSweep.txt};
            \addlegendentry{$\theta = 0.60$}
         \end{axis}
        \end{tikzpicture}
    }
    \hspace{-0.42cm}
   \subfloat{
   \raisebox{0em}{
    \begin{tikzpicture}[baseline]
        \begin{axis}[
            ybar=1pt,
    legend image code/.code={
        \draw [#1] (0cm,-0.15cm) rectangle (5pt,0.15cm);
    },
            width=6.5cm,
            height=5.28cm,
            bar width=3pt,
            xtick pos=bottom,
            ytick pos=left,
            xticklabel style={font=\footnotesize,            
                yshift=1ex,
                xshift=5pt,
                rotate=45,
                anchor=east
            },
            ytick={0,10,20,30,40},
            yticklabel style={rotate=90, font=\footnotesize}, 
            ylabel={\footnotesize {Latency Reduction (\%)}},
            xlabel={\footnotesize{SNR (dB)}\\[-0.5ex] \footnotesize{(c)}}, 
                xlabel style={yshift=4mm, align=center},
            ylabel style={yshift=-1.5mm},
            symbolic x coords=
            {$-1.0$, $-0.6$, $-0.1$, 0.5, 1.3, 2.3, 2.8, 3.3},
            xtick={$-1.0$, $-0.6$, $-0.1$, 0.5, 1.3, 2.3, 2.8, 3.3},
            enlarge x limits=0.15,
            nodes near coords={\pgfmathprintnumber\pgfplotspointmeta\%},
            every node near coord/.append style={font=\tiny, color=black, rotate=90, anchor=west, yshift=0pt},
            ymin=0, ymax=52,
            clip=false,
            grid=major,
            grid style={draw=lightgrid},
            tick style={draw=none},
            legend style={
                row sep=0.02cm,
                font=\footnotesize,
                cells={align=left},
                legend cell align={left},
                nodes={scale=0.85, transform shape},
                at={(0.02,0.98)},
                anchor=north west
            },
        ]
            
            \addplot[bar shift=-4.5pt,fill=myblue!80, draw=myblue!30!black, thick] coordinates {
                ($-1.0$, 6.6)
                ($-0.6$, 11.6)
                ($-0.1$, 14.7)
                (0.5, 19.1)
                (1.3, 25.8)
                (2.3, 13.9)
                (2.8, 33.1)
                (3.3, 28.7)
                };
            \addlegendentry{$\theta = 0.20$}
            
            \addplot[bar shift= 0pt,fill=myred!70, draw=myred!30!black, thick] coordinates {
                ($-1.0$, 10.1)
                ($-0.6$, 15.1)
                ($-0.1$, 18.7)
                (0.5, 20.6)
                (1.3, 26.3)
                (2.3, 25.5)
                (2.8, 35.7)
                (3.3, 28.5)
            };
            \addlegendentry{$\theta = 0.40$}

            \addplot[bar shift=4.5pt,fill=gray!60, draw=gray!30!black, thick] coordinates {
                ($-1.0$, 9.4) ($-0.6$, 16.3) ($-0.1$, 20.2)
                (0.5, 22.3) (1.3, 26.8) (2.3, 21.7)
                (2.8, 33.9) (3.3, 29.2)
            };
            \addlegendentry{$\theta = 0.60$}
            \end{axis}
        \end{tikzpicture}
    }}
    \vspace*{-0.3cm}
    \caption{Impact of the threshold $\theta$ on decoding performance under the baseline schedule for BG1 with $A{=}4200$ and $R{=}1/3$. (a) BLER {\emph{vs.}} SNR, highlighting the emergence of error floors at larger $\theta$ values. (b) Average number of processed layers and (c) corresponding percentage of latency reduction.}
    \label{thetasweep}
    \vspace*{-0.4cm}
\end{figure*}

%% file: evalplot.tex
\begin{figure*}[t]
        \centering
        \parbox{\textwidth}{
        \hspace{-0.4cm}
        \subfloat{
            \begin{tikzpicture}[baseline=(current axis.south)]
                \begin{semilogyaxis}[
                    width=6.5cm, height=5.3cm,
                    xmin=-1.2, xmax=-0.5, ymin=1e-5, ymax=1,
                    xlabel={\footnotesize{SNR (dB)} \\[-0.75ex] \footnotesize{(a)}}, 
                    xlabel style={yshift=1mm, align=center},
                    ylabel={\footnotesize {BLER}}, ylabel style={yshift=-1mm},
                    x tick label style={font=\footnotesize, yshift=0mm},
                    y tick label style={ font=\footnotesize}, 
                    grid=both, minor tick num=1,
                    legend pos=south west,
                    legend style={
                    row sep=0.02cm,
                    font=\footnotesize,
                    cells={align=left},
                    legend cell align={left},
                    nodes={scale=0.75, transform shape},
                    at={(0.001,0.001)}
                },]
                \addplot[myred!70, solid, mark=square, line width=1.1pt] table[x=SNR, y=Letters]{Letters_BLER_Data_A3848.txt}; \addlegendentry{BGSS~\cite{9765969}}
                \addplot[myred!70, dashed, mark=square*, line width=1.2pt] table[x=SNR, y=Letters_CoreLock]{Letters_BLER_Data_A3848.txt}; \addlegendentry{BGSS~\cite{9765969}${+}$Core-lock}
                
                \addplot[myblue!80, solid, mark=diamond, line width=1.1pt] table[x=SNR, y=BtT]{BtT_BLER_Data_A3848.txt}; \addlegendentry{BtT~\cite{8661345}}
                \addplot[myblue!80, dashed, mark=diamond*, line width=1.2pt] table[x=SNR, y=BtT_CoreLock]{BtT_BLER_Data_A3848.txt}; \addlegendentry{BtT~\cite{8661345}${+}$Core-lock}
                
                \end{semilogyaxis}
            \end{tikzpicture}
        }
        \hspace{-0.3cm}
        \subfloat{
    \begin{tikzpicture}[baseline=(plotBot.south)]
        
        \begin{axis}[
            name=plotTop, 
            width=6.5cm, height=3.25cm, 
            xmin=-1.2, xmax=-0.4, 
            ymin=4.7, ymax=15.3,  
            ylabel style={yshift=-1mm},
            x tick label style={font=\footnotesize},
            y tick label style={font=\footnotesize}, 
            grid=both, minor tick num=1,
            legend style={
                row sep=0.02cm, font=\scriptsize, cells={align=left},
                legend cell align={left}, nodes={scale=0.5, transform shape},
                at={(0.98,0.98)}, anchor=north east
            }
        ]
        \addplot[myred!70, solid, mark=square, mark size=1pt, line width=1.1pt] table[x=SNR, y=Letters_ANI]{Letters_ANI_Data_A3848.txt};
        \addlegendentry{BGSS~\cite{9765969}}

        \addplot[myred!70, dashed, mark=square*, mark size=1.2pt, line width=1.2pt] table[x=SNR, y=Letters_CoreLock_ANI]{Letters_ANI_Data_A3848.txt};
        \addlegendentry{BGSS~\cite{9765969}${+}$Core-lock}
        
        \addplot[myblue!80, solid, mark=diamond, mark size=1.2pt, line width=1.2pt] table[x=SNR, y=BtT_ANI]{BtT_ANI_Data_A3848.txt};
        \addlegendentry{BtT~\cite{8661345}}
        \addplot[myblue!80, dashed, mark=diamond*, mark size=1.2pt, line width=1.2pt] table[x=SNR, y=BtT_CoreLock_ANI]{BtT_ANI_Data_A3848.txt};
       \addlegendentry{BtT~\cite{8661345}${+}$Core-lock}
        \end{axis}
        
        \begin{axis}[
            name=plotBot,
            at={(plotTop.south)}, anchor=north, yshift=-0.4cm, 
            width=6.5cm, height=3.25cm, 
            xmin=-0.3, xmax=2.5, 
            ymin=0.9, ymax=5, 
            xlabel={\footnotesize{SNR (dB)}\\[-0.75ex] \footnotesize{(b)}}, 
            xlabel style={yshift=1mm, align=center},
            x tick label style={font=\footnotesize},
            y tick label style={font=\footnotesize}, 
            grid=both, minor tick num=1
        ]
        \addplot[myred!70, solid, mark=square, mark size=1pt, line width=1.1pt] table[x=SNR, y=Letters_ANI]{Letters_ANI_Data_A3848.txt};

        \addplot[myred!70, dashed, mark=diamond*, mark size=1.2pt, line width=1.2pt] table[x=SNR, y=Letters_CoreLock_ANI]{Letters_ANI_Data_A3848.txt};
        
        \addplot[myblue!80, solid, mark=diamond*, mark size=1.2pt, line width=1.2pt] table[x=SNR, y=BtT_ANI]{BtT_ANI_Data_A3848.txt};

        \addplot[myblue!80, dashed, mark=diamond*, mark size=1.2pt, line width=1.2pt] table[x=SNR, y=BtT_CoreLock_ANI]{BtT_ANI_Data_A3848.txt};

        \end{axis}
        \path (plotTop.north west) -- (plotBot.south west) 
            node[midway, xshift=-0.5cm, rotate=90, font=\footnotesize] {Average Number of Iterations};
    \end{tikzpicture}
}
        \hspace{-0.5cm}
        \subfloat{
        \begin{tikzpicture}[baseline=(current axis.south)]
            \begin{axis}[
                ybar=1pt,
                legend image code/.code={\draw [#1] (0cm,-0.15cm) rectangle (5pt,0.15cm);},
                width=6.5cm, 
                height=5.3cm,
                bar width=5pt,
                xtick pos=bottom, ytick pos=left,
                xticklabel style={font=\footnotesize, yshift=1ex, xshift=5pt, rotate=45, anchor=east},
                ytick={0,10,20,30,40},
                yticklabel style={rotate=90, font=\footnotesize}, 
                ylabel={\footnotesize{Latency Reduction (\%)}},
                xlabel={\footnotesize{SNR (dB)} \\[-0.75ex] \footnotesize{(c)}}, 
                xlabel style={yshift=4mm, align=center},
                symbolic x coords={$-0.8$, $-0.5$, $-0.2$, 0.1, 0.9, 1.7, 2.5},
                xtick={$-0.8$, $-0.5$, $-0.2$, 0.1, 0.9, 1.7, 2.5},
                enlarge x limits=0.12,
                nodes near coords={\pgfmathprintnumber\pgfplotspointmeta\%},
                every node near coord/.append style={font=\tiny, color=black, rotate=90, anchor=west, yshift=0pt},
                ymin=0, ymax=45,
                clip=false, grid=major, grid style={draw=lightgray, dashed},
                tick style={draw=none},
                legend style={
                    row sep=0.02cm, font=\footnotesize, cells={align=left},
                    legend cell align={left}, nodes={scale=0.85, transform shape},
                    at={(0.02,0.98)}, anchor=north west
                }
            ]
            
            \addplot[fill=myred!70, draw=myred!30!black, thick] coordinates {
                ($-0.8$, 10.7)
                ($-0.5$, 14.2)
                ($-0.2$, 16.0)
                (0.1, 17.7)
                (0.9, 17.7)
                (1.7, 29.1)
                (2.5, 20.3)
            };
            \addlegendentry{Core-lock gain in~\cite{9765969}}
            
            \addplot[fill=myblue!80, bar shift=4pt, draw=myblue!30!black, thick] coordinates {
                ($-0.8$, 10.8)
                ($-0.5$, 14.3)
                ($-0.2$, 16.0)
                (0.1, 18.5)
                (0.9, 19.7)
                (1.7, 31.3)
                (2.5, 18.3)
            };
            \addlegendentry{Core-lock gain in~\cite{8661345}}

            \end{axis}
        \end{tikzpicture}
        }\\[-0.43cm]
    \hspace*{-0.4cm}
     \centering
        \subfloat{
            \begin{tikzpicture}[baseline=(current axis.south)]
                \begin{semilogyaxis}[
                    width=6.5cm, height=5.3cm,
                    xmin=-1.2, xmax=-0.6, ymin=1e-5, ymax=1,
                    xlabel={\footnotesize{SNR (dB)} \\[-0.75ex] \footnotesize{(d)}}, 
                    xlabel style={yshift=1mm, align=center},
                    ylabel={\footnotesize {BLER}}, ylabel style={yshift=-1mm},
                    x tick label style={font=\footnotesize, yshift=0mm},
                    y tick label style={ font=\footnotesize}, 
                    grid=both, minor tick num=1,
                    legend pos=south west,
                    legend style={
                    row sep=0.02cm,
                    font=\footnotesize,
                    cells={align=left},
                    legend cell align={left},
                    nodes={scale=0.75, transform shape},
                    at={(0.001,0.001)}
                },]
                \addplot[mygray!60, solid, mark=square, line width=1.1pt] table[x=SNR, y=BellLabs]{BellLabs_BLER_Data_A3848.txt}; \addlegendentry{EO/EC/ECC/CO~\cite{11154636}}
                \addplot[mygray!60, dashed, mark=square*, line width=1.2pt] table[x=SNR, y=BellLabs_CoreLock]{BellLabs_BLER_Data_A3848.txt}; \addlegendentry{EO/EC/ECC/CO~\cite{11154636}${+}$Core-lock}

                \addplot[mypurple, solid, mark=triangle, line width=1.1pt] table[x=SNR, y=BellLabs_no_EO]{BellLabs_no_EO_BLER_Data_A3848.txt}; \addlegendentry{$\neq$EO${+}$EC/ECC/CO}
                \addplot[mypurple, dashed, mark=triangle*, line width=1.2pt] table[x=SNR, y=BellLabs_no_EO_CoreLock]{BellLabs_no_EO_BLER_Data_A3848.txt}; \addlegendentry{$\neq$EO${+}$EC/ECC/CO${+}$Core-lock}
                
                \end{semilogyaxis}
            \end{tikzpicture}
        }
        \hspace{-0.6cm}
        \subfloat{
    \begin{tikzpicture}[baseline=(plotBot.south)]
        
        \begin{axis}[
            name=plotTop, 
            width=6.5cm, height=3.25cm, 
            xmin=-1, xmax=-0.4, 
            ymin=4.7, ymax=12.5,  
            ylabel style={yshift=-1mm},
            x tick label style={font=\footnotesize},
            y tick label style={font=\footnotesize}, 
            grid=both, minor tick num=1,
            legend style={
                row sep=0.02cm, font=\scriptsize, cells={align=left},
                legend cell align={left}, nodes={scale=0.52, transform shape},
                at={(0.98,0.98)}, anchor=north east
            }
        ]
        \addplot[mygray!60, solid, mark=square, mark size=1pt, line width=1.1pt] table[x=SNR, y=BellLabs_ANI]{BellLabs_ANI_Data_A3848.txt};
        \addlegendentry{EO/EC/ECC/CO~\cite{11154636}}
        
        \addplot[mygray!60, dashed, mark=square*, mark size=1.2pt, line width=1.2pt] table[x=SNR, y=BellLabs_CoreLock_ANI]{BellLabs_ANI_Data_A3848.txt};
        \addlegendentry{EO/EC/ECC/CO~\cite{11154636}${+}$Core-lock}
        

        \addplot[mypurple, dashed, mark=triangle*, mark size=1.2pt, line width=1.2pt] table[x=SNR, y=BellLabs_no_EO_CoreLock_ANI]{BellLabs_no_EO_ANI_Data_A3848.txt};
        \addlegendentry{$\neq$EO${+}$EC/ECC/CO${+}$Core-lock}
        \end{axis}
        
        \begin{axis}[
            name=plotBot,
            at={(plotTop.south)}, anchor=north, yshift=-0.4cm, 
            width=6.5cm, height=3.25cm, 
            xmin=-0.3, xmax=2.5, 
            ymin=0.9, ymax=5, 
            xlabel={\footnotesize{SNR (dB)}\\[-0.75ex] \footnotesize{(e)}}, 
            xlabel style={yshift=1mm, align=center},
            x tick label style={font=\footnotesize},
            y tick label style={font=\footnotesize}, 
            grid=both, minor tick num=1
        ]
        \addplot[mygray!60, solid, mark=square, mark size=1pt, line width=1.1pt] table[x=SNR, y=BellLabs_ANI]{BellLabs_ANI_Data_A3848.txt};

        \addplot[mygray!60, dashed, mark=diamond*, mark size=1.2pt, line width=1.2pt] table[x=SNR, y=BellLabs_CoreLock_ANI]{BellLabs_ANI_Data_A3848.txt};
        

        \addplot[mypurple, dashed, mark=triangle*, mark size=1.2pt, line width=1.2pt] table[x=SNR, y=BellLabs_no_EO_CoreLock_ANI]{BellLabs_no_EO_ANI_Data_A3848.txt};
        \end{axis}
        \path (plotTop.north west) -- (plotBot.south west) 
            node[midway, xshift=-0.5cm, rotate=90, font=\footnotesize] {Average Number of Iterations};
    \end{tikzpicture}
}
        \hspace{-0.5cm}
        \subfloat{
        \begin{tikzpicture}[baseline=(current axis.south)]
            \begin{axis}[
                ybar=1pt,
                legend image code/.code={\draw [#1] (0cm,-0.15cm) rectangle (5pt,0.15cm);},
                width=6.5cm, height=5.3cm,
                bar width=7pt,
                xtick pos=bottom, ytick pos=left,
                xticklabel style={font=\footnotesize, yshift=1ex, xshift=5pt, rotate=45, anchor=east},
                ytick={0,10,20,30,40,50,60},
                yticklabel style={rotate=90, font=\footnotesize}, 
                ylabel={\footnotesize{Latency Reduction (\%)}},
                xlabel={\footnotesize{SNR (dB)} \\[-0.75ex] \footnotesize{(f)}}, 
                xlabel style={yshift=4mm, align=center},
                ylabel style={yshift=-1mm},
                symbolic x coords={$-0.8$, $-0.6$, $-0.4$, $-0.2$, 0, 0.1, 0.5, 0.9, 1.7, 2.5},
                xtick={$-0.8$, $-0.6$, $-0.4$, $-0.2$, 0, 0.1, 0.5, 0.9, 1.7, 2.5},
                enlarge x limits=0.12,
                nodes near coords={\pgfmathprintnumber\pgfplotspointmeta\%},
                every node near coord/.append style={font=\tiny, color=black, rotate=90, anchor=west, yshift=0pt},
                ymin=0, ymax=60,
                clip=false, grid=major, grid style={draw=lightgray, dashed},
                tick style={draw=none},
                legend style={
                    row sep=0.02cm, font=\footnotesize, cells={align=left},
                    legend cell align={left}, nodes={scale=0.6, transform shape},
                    at={(0.02,0.98)}, anchor=north west
                }
            ]
            
          \addplot[fill=mypurple, draw=mypurple!30!black, thick] coordinates {
                ($-0.8$, 4.6)
                ($-0.6$, 3.8)
                ($-0.4$, 4.6)
                ($-0.2$, 8.7)
                (0, 8.2)
                (0.1, 14.4)
                (0.5, 13.8)
                (0.9, 28.1)
                (1.7, 12.4)
                (2.5, 45.5)
            };
            \addlegendentry{$\neq$EO${+}$EC/ECC/CO${+}$Core-lock over~\cite{11154636}}

            \end{axis}
        \end{tikzpicture}
        }
        \vspace*{-0.3cm}
    \caption{Core-lock evaluation in static schedules (a)--(c) and dynamic multi-state architectures (d)--(f) for BG1, $A{=}3848$ and $R{=}1/3$. Subplots (a) and (d)  depict BLER \emph{vs.} SNR, (b) and (e) show absolute average number of iterations, while (c) and (f) show the corresponding percentage of latency reduction.}
    \label{maineval}
    \vspace*{-0.4cm}
    }
\end{figure*}

%% file: FigArch.tex
\begin{tikzpicture}
    \node [rectangle, 
    cyan!80!black,
    minimum height=0.8cm,
    minimum width=1.2cm,
    inner sep=2pt, 
    outer sep=0pt,
    fill=cyan!10!white,draw] (XORTREE) at (5,4.25*0.75) {
    \centering
    \begin{tikzpicture}[circuit logic US, 
    circuit symbol size=width 1 height 1,
    circuit symbol unit=2pt,scale=0.75]
    \foreach \x/\y/\ll in 
        {1/1/1,1/2/2,2/1.5/3,
        1/3/4,1/4/5,2/3.5/6,
        3/2.5/7} {
    \node [xor gate,fill=cyan,draw=cyan!75!black] (x\ll) at (\x*0.75,\y*0.5) {}; 
    }
    \draw [cyan!75!black] (x1.output) -- ++(right:1mm) |- (x3.input 2);
    \draw [cyan!75!black](x2.output) -- ++(right:1mm) |- (x3.input 1);
    \draw [cyan!75!black](x4.output) -- ++(right:1mm) |- (x6.input 2);
    \draw [cyan!75!black](x5.output) -- ++(right:1mm) |- (x6.input 1);
    \draw [cyan!75!black](x3.output) -- ++(right:1mm) |- (x7.input 2);
    \draw [cyan!75!black](x6.output) -- ++(right:1mm) |- (x7.input 1);
    \end{tikzpicture}
    };
    \foreach \y in {1,2,3,4}{
    \node [draw=blue!80!black, 
        fill=blue!20!white,
        rectangle, rounded corners,
        text width=1.5cm, 
        font=\footnotesize,
        inner sep=1pt] (BR\y) at (0,\y*0.75) {BR 384 bits 19 inst.} ; 
    \node [draw=green!80!black,
        fill=green!20!white, 
        rectangle, rounded corners,
        text width=1.25cm, 
        font=\footnotesize,
        inner sep=1pt] (XOR\y) at (2.6,\y*0.75) {XOR trees 384 inst.} ; 
    \draw [-latex] (BR\y) -> (XOR\y) node [pos=0.45,fill=white,inner sep=0pt,font=\tiny]{$384{\times}19$};
    }
    \node [draw=orange!80!black,
        fill=orange!20!white,
        rectangle, rounded corners,
        inner sep=1pt, 
        font=\footnotesize,
        text width=1.1cm,
        minimum height=0.75cm] (ADD) at (4.75,2.5*0.75) {Adder 1536 bits};
    \foreach \y/\a/\ll in {1/205/3,2/190/2,3/170/2,4/155/3}
    {
     \draw [-latex] (XOR\y.east) -- ++(right:\ll*1mm) |- (ADD.\a)   
        node [pos=0.7,font=\tiny,fill=white,inner sep=0pt] {$384$};
    }
    \node [draw] (Comp) at (5.9,2.5*0.75) {$>$};
    \draw [-latex](ADD)--(Comp);
    \node [draw,below of=Comp,
        font=\footnotesize] (ROM) {ROM $Z_c \cdot \theta$};
    \draw [-latex] (ROM)--(Comp);

    \node [draw=teal!80!black, 
        rectangle, rounded corners,
        right of=Comp,
        text width=1.2cm,
        inner sep=1pt,
        node distance=1.15cm,
        font=\footnotesize,
        fill=teal!20!white
        ] 
        (IND) 
        {Index Extraction};
    \draw [-latex] (Comp)--(IND);

    \drawarrow[bend left]{XOR4.110}{XORTREE.145}
   
    \end{tikzpicture}

%% file: timing.tex
\begin{figure}[t]
\centering
\begin{tikzpicture}[
    scale=1.0,
    layer/.style={draw, minimum width=0.42cm, minimum height=0.45cm, align=center, font=\scriptsize\bfseries, inner sep=1pt},
    arrow/.style={->, >=Stealth, thick}
]

\foreach \step/\x/\color in {
    A/0/myblue!20, B/1/myblue!20, C/2/myblue!20, D/3/myblue!20,
    E/4.3/myblue!20, F/5.3/myblue!20, G/6.3/myblue!20, H/7.3/myblue!20,
    I/8.6/myblue!20, J/9.6/mygreen!20, K/10.6/mygreen!20, L/11.6/mygreen!20,
    M/12.9/mygreen!20, N/13.9/mygreen!20, O/14.9/mygreen!20, P/15.9/mygreen!20
} {
    \node[layer, fill=\color] (\step) at (\x*0.51, 0) {\step};
}

\node[above=0.03cm of A, font=\tiny] {CS};
\node[above=0.03cm of D, font=\tiny] {CE};
\node[above=0.03cm of E, font=\tiny] {CS};
\node[above=0.03cm of H, font=\tiny] {CE};
\node[above=0.03cm of I, font=\tiny] {CS};
\node[above=0.03cm of L, font=\tiny] {CE};
\node[above=0.03cm of M, font=\tiny] {CS};
\node[above=0.03cm of P, font=\tiny, text=myred] {CE};

\draw[arrow, mygreen!80!black, thick] (A.south) -- ++(0, -0.45) 
    node[below, 
    font=\tiny\bfseries,
    text=mygreen!80!black,
    yshift=0.02cm,
    ] {FPC};

\node[layer, fill=mygreen!20] (B_prime_parallel) at ($(B.south)+(0, -0.6)$) {B$'$};
\node[below=0.02cm of B_prime_parallel, 
yshift=0.1cm,
font=\tiny\bfseries, text=mygreen!80!black] {ACL};

\draw[arrow, mygreen!80!black, thick] (B.south) -- (B_prime_parallel.north);

\draw[arrow, myred, thick] ($(B_prime_parallel.east)+(0.3, 0.3)$) -- (B_prime_parallel.east) node[pos=0.1, 
anchor=west,
xshift=-0.1cm,
align=center, font=\bfseries\tiny, text=myred] {FPC End\\(Codeword Found)};

\node[layer, fill=mygreen!20] (B_prime_serial) at ($(I.south)+(0, -0.6)$) {B$'$};
\node[below=0.01cm of B_prime_serial, 
yshift=0.1cm,
font=\tiny\bfseries,
text=mygreen!80!black] {ACL};

\draw[arrow, mygreen!80!black, thick] (I.south) -- (B_prime_serial.north);

\draw[arrow, myred, thick] ($(P.south east)+(-0.3, -0.3)$) -- (P.south east);
\node[left=0.05cm of {$(P.south east)+(-0.1, -0.5)$}, font=\bfseries\tiny, text=myred, align=right] {End\\(Codeword Found)};

\draw [-latex] (B_prime_parallel.south east) ++(right: 0.5cm) -- ++(right: 2cm) node[right,
xshift=-0.1cm,
font=\scriptsize\bfseries, text=black] {Time};


\end{tikzpicture}
\vspace{-0.3cm}
\caption{Core-lock evaluation timeline for fully parallel and multiplexed execution. Letters A--P denote sub-iteration layers, CS is Check Start, CE is  Check End, FPC is  Fully Parallel Check, and ACL is Apply Core-Lock.}
\label{fig:delay}
\vspace{-0.3cm}
\end{figure}
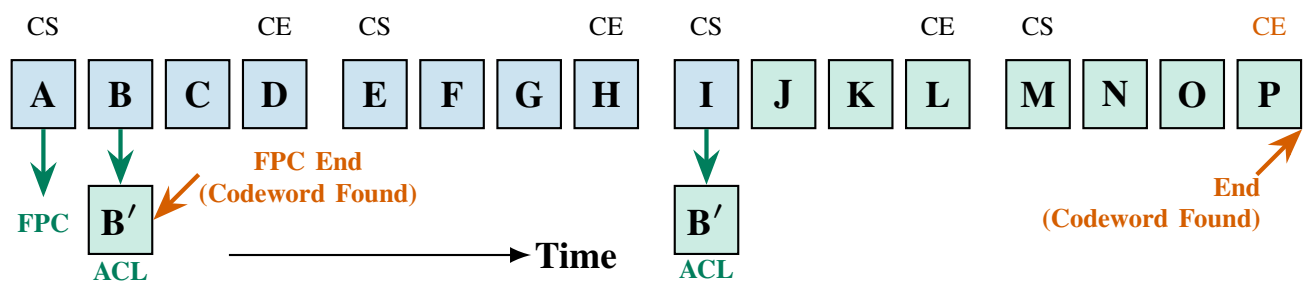

%% file: main_arxiv.bbl
\begin{thebibliography}{10}
\providecommand{\url}[1]{#1}
\csname url@samestyle\endcsname
\providecommand{\newblock}{\relax}
\providecommand{\bibinfo}[2]{#2}
\providecommand{\BIBentrySTDinterwordspacing}{\spaceskip=0pt\relax}
\providecommand{\BIBentryALTinterwordstretchfactor}{4}
\providecommand{\BIBentryALTinterwordspacing}{\spaceskip=\fontdimen2\font plus
\BIBentryALTinterwordstretchfactor\fontdimen3\font minus \fontdimen4\font\relax}
\providecommand{\BIBforeignlanguage}[2]{{%
\expandafter\ifx\csname l@#1\endcsname\relax
\typeout{** WARNING: IEEEtran.bst: No hyphenation pattern has been}%
\typeout{** loaded for the language `#1'. Using the pattern for}%
\typeout{** the default language instead.}%
\else
\language=\csname l@#1\endcsname
\fi
#2}}
\providecommand{\BIBdecl}{\relax}
\BIBdecl

\bibitem{9711564}
{Vaezi et al.}, ``{Cellular, Wide-Area, and Non-Terrestrial IoT: A Survey on 5G Advances and the Road Toward 6G},'' \emph{IEEE Communications Surveys \& Tutorials}, vol.~24, no.~2, pp. 1117--1174, 2022.

\bibitem{10975810}
{Tuninato et al.}, ``{5G New Radio for Non-Terrestrial Networks: Analysis and Comparison of HARQ and RLC ARQ Performance Over Satellite Links},'' \emph{IEEE Access}, vol.~13, pp. 75\,400--75\,415, 2025.

\bibitem{1057683}
R.~Gallager, ``{Low-density parity-check codes},'' \emph{IRE Transactions on Information Theory}, vol.~8, no.~1, pp. 21--28, 1962.

\bibitem{etsi_ts_138212_v17_5_0}
``{5G; NR; Multiplexing and channel coding (3GPP TS 38.212 version 17.5.0 Release 17)},'' European Telecommunications Standards Institute (ETSI), Tech. Rep. TS 138 212 V17.5.0, Apr. 2023, superseded by ETSI TS 138 212 V17.6.0:2023-10.

\bibitem{910572}
F.~Kschischang, B.~Frey, and H.-A. Loeliger, ``{Factor graphs and the sum-product algorithm},'' \emph{IEEE Transactions on Information Theory}, vol.~47, no.~2, pp. 498--519, 2001.

\bibitem{768759}
{Fossorier et al.}, ``{Reduced complexity iterative decoding of low-density parity check codes based on belief propagation},'' \emph{IEEE Transactions on Communications}, vol.~47, no.~5, pp. 673--680, 1999.

\bibitem{1001666}
J.~Chen and M.~Fossorier, ``{Density evolution for two improved BP-Based decoding algorithms of LDPC codes},'' \emph{IEEE Communications Letters}, vol.~6, no.~5, pp. 208--210, 2002.

\bibitem{5610969}
A.~I.~V. Casado, M.~Griot, and R.~D. Wesel, ``{LDPC Decoders with Informed Dynamic Scheduling},'' \emph{IEEE Transactions on Communications}, vol.~58, no.~12, pp. 3470--3479, 2010.

\bibitem{8664167}
X.~Liu, L.~Zi, D.~Yang, and Z.~Wang, ``{Improved Decoding Algorithms of LDPC Codes Based on Reliability Metrics of Variable Nodes},'' \emph{IEEE Access}, vol.~7, pp. 35\,769--35\,778, 2019.

\bibitem{8661345}
J.~Frenzel, S.~Mueller-Weinfurtner, J.~B. Huber, and R.~R. Mueller, ``{Static Layered Schedules and Core-Only Parity Check for the 5G New Radio LDPC Codes},'' in \emph{SCC 2019; 12th International ITG Conference on Systems, Communications and Coding}, 2019, pp. 1--5.

\bibitem{9765969}
K.~Tian and H.~Wang, ``{A Novel Base Graph Based Static Scheduling Scheme for Layered Decoding of 5G LDPC Codes},'' \emph{IEEE Communications Letters}, vol.~26, no.~7, pp. 1450--1453, 2022.

\bibitem{papageorgiou2025novel}
{N. Papageorgiou et al.}, ``{Novel Scheduling and Shifter Networks for 5G LDPC Decoders},'' in \emph{International Workshop on Design and Architectures for Signal and Image Processing}.\hskip 1em plus 0.5em minus 0.4em\relax Springer, 2025, pp. 95--107.

\bibitem{9222014}
S.~Shao, Y.~Zhang, R.~G. Maunder, and L.~Hanzo, ``{3D EXIT Charts for Analyzing the 5G 3GPP New Radio LDPC Decoder},'' \emph{IEEE Access}, vol.~8, pp. 188\,797--188\,812, 2020.

\bibitem{11154636}
A.~Elkelesh and J.~Ling, ``{Optimizing Schedulers for Layered BP Decoding: From Code-Specific to Universal Decoding},'' in \emph{2025 13th International Symposium on Topics in Coding (ISTC)}, 2025, pp. 1--5.

\bibitem{1363033}
D.~Hocevar, ``{A reduced complexity decoder architecture via layered decoding of LDPC codes},'' in \emph{IEEE Workshop on Signal Processing Systems, 2004. SIPS 2004.}, 2004, pp. 107--112.

\bibitem{8316763}
T.~Richardson and S.~Kudekar, ``{Design of Low-Density Parity Check Codes for 5G New Radio},'' \emph{IEEE Communications Magazine}, vol.~56, no.~3, pp. 28--34, 2018.

\bibitem{8052499}
I.~Tsatsaragkos and V.~Paliouras, ``A reconfigurable ldpc decoder optimized for 802.11n/ac applications,'' \emph{IEEE Transactions on Very Large Scale Integration (VLSI) Systems}, vol.~26, no.~1, pp. 182--195, 2018.

\end{thebibliography}
